# Keyword-aware Optimal Route Search

Xin Cao    Lisi Chen    Gao Cong    Xiaokui Xiao

School of Computer Engineering, Nanyang Technological University, Singapore

{xcao1,lchen012}@e.ntu.edu.sg,{gaocong,xkxiao}@ntu.edu.sg

## ABSTRACT

Identifying a preferable route is an important problem that finds applications in map services. When a user plans a trip within a city, the user may want to find "a most popular route such that it passes by *shopping mall*, *restaurant*, and *pub*, and the travel time to and from his hotel is within 4 hours." However, none of the algorithms in the existing work on route planning can be used to answer such queries. Motivated by this, we define the problem of keyword-aware optimal route query, denoted by KOR, which is to find an optimal route such that it covers a set of user-specified keywords, a specified budget constraint is satisfied, and an objective score of the route is optimal. The problem of answering KOR queries is NP-hard. We devise an approximation algorithm OSScaling with provable approximation bounds. Based on this algorithm, another more efficient approximation algorithm BucketBound is proposed. We also design a greedy approximation algorithm. Results of empirical studies show that all the proposed algorithms are capable of answering KOR queries efficiently, while the BucketBound and Greedy algorithms run faster. The empirical studies also offer insight into the accuracy of the proposed algorithms.

## 1. INTRODUCTION

Identifying a preferable route in a road network is an important problem that finds applications in map services. For example, map applications like Baidu Lvyou [1] and Yahoo Travel [2] offer tools for trip planning. However, the routes that they provide are collected from users and are thus pre-defined. This is a significant deficiency since there may not exist any pre-defined route that meets the user needs. The existing solutions (e.g., [16, 17, 22]) for trip planning or route search are often insufficient in offering the flexibility for users to specify their requirements on the route.

Consider a user who wants to spend a day exploring a city. She is not familiar with the city and she might pose such a query: "Find the most popular route to and from my hotel such that it passes by *shopping mall*, *restaurant*, and *pub*, and the time spent on the road in total is within 4 hours."

---
[1] http://lvyou.baidu.com/
[2] http://travel.yahoo.com

The example query above has two hard constraints: 1) the points of interests preferred by the user, as expressed by a set of keywords that should be covered in the route (e.g., "shopping mall", "restaurant" and "pub"); 2) a budget constraint (e.g., travel time) that should be satisfied by the route. The query aims to identify the optimal route under the two hard constraints, such that an objective score is optimized (e.g., route popularity [4]). Note that route popularity can be estimated by the number of users traveling a route, obtained from the user traveling histories recorded in sources such as GPS trajectories or Flickr photos [4]. In general, the budget constraint and the objective score can be of various different types, such as travel duration, distance, popularity, travel budget, etc. We consider two different attributes for budget constraint and objective score because users often need to balance the trade-off of two aspects when planning their trips. For example, a popular route may be quite expensive, or a route with the shortest length is of little interests. In the example query, it is likely that the most popular route requires traveling time more than 4 hours. Hence, a route searching system should be able to balance such trade-offs according to users' different preferences.

We refer to the aforementioned type of queries as *keyword-aware optimal route query*, denoted as KOR. Formally, a KOR query is defined over a graph $\mathcal{G}$, and the input to the query consists of five parameters, $v_s, v_t, \psi, \Delta$, and $f$, where $v_s$ is the source location of the route in $G$, $v_t$ is the target location, $\psi$ is a set of keywords, $\Delta$ is a budget limit, and $f$ is a function that calculates the objective score of a route. The query returns a path $R$ in $G$ starting at $v_s$ and ending at $v_t$, such that $R$ minimizes $f(R)$ under the constraints that $R$ satisfies the budget limit $\Delta$ and passes through locations that cover the query keywords in $\psi$. To the best of our knowledge, none of the existing work on trip planning or route search (e.g., [16, 17, 22]) is applicable for KOR queries. Furthermore, the problem of solving KOR queries can be shown to be NP-hard by a reduction from the weighted constrained shortest path problem [8]. It can also be viewed as a generalized traveling salesman problem [11] with constraints. This leads to an interesting question: is it possible to derive efficient solutions to answering KOR queries?

Due to the hardness of answering KOR queries, in this paper, we answer the aforementioned question affirmatively with three approximation algorithms. The first approximation algorithm has a performance bound and is denoted by OSScaling. In OSScaling, we first scale the objective value of every edge to an integer by a parameter $\epsilon$ to obtain a scaled graph denoted by $\mathcal{G}_S$. Specifically, in the scaled graph $\mathcal{G}_S$, each partial route is represented by a "label", which records the query keywords already covered by the partial route, the scaled objective score, the original objective score, and the budget score of the route. At each node, we maintain a list of "useful" labels corresponding to the routes that go to that node.

Starting from the source node, we keep creating new partial routes by extending the current "best" partial route to generate new labels, until all the potentially useful labels on the target node are generated. Finally, the route represented by the label with the best objective score at the target node is returned.

We prove that the algorithm returns routes with objective scores no worse than $\frac{1}{1-\epsilon}$ times of that of the optimal route. The worst case complexity of OSScaling is polynomial with $\frac{1}{\epsilon}$, the budget constraint $\Delta$, the number of edges and nodes in $\mathcal{G}$, and it is exponential in the number of query keywords, which is usually small in our targeted applications, as it is well known that search engine queries are short, and an analysis on a large Map query log [25] shows that nearly all queries contain fewer than 5 words.

Our second algorithm improves on the algorithm OSScaling, which is referred to as BucketBound. It also returns approximate solutions to KOR queries with performance guarantees. However, it is more efficient than OSScaling. The algorithm can always return a route whose objective score is at most $\beta$ ($\beta > 1$ is a parameter) times of the one found by OSScaling. The algorithm divides the traversed partial routes into different "buckets" according to the best possible objective scores they can achieve. This enables us to develop a novel way to detect if a feasible route (covering all query keywords and satisfying the budget constraint) is in the same bucket with the one found by OSScaling. When we find a feasible route that falls in the same bucket as the route found by OSScaling, we return it as the result.

Finally, we also present a greedy approach for the problem. From the starting location, we keep selecting the next location greedily, taking into account all the three constraints in the KOR query. This is repeated until we reach the target location. This algorithm is efficient, although it may generate a route that violates the two hard constraints of KOR: covering all query keywords and satisfying the budget constraint.

In summary, our contributions are threefold. First, we propose the keyword-aware optimal route (KOR) query, and we show that the problem of solving KOR queries is NP-hard. Second, we present two novel approximation algorithms both with provable performance bounds for the KOR problem. We also provide a greedy approach. Third, we study the properties of the paper's proposals empirically on a graph extracted from a large collection of Flickr photos. The results demonstrate that the proposed solutions offer scalability and excellent performance.

The rest of the paper is organized as follows: Section 2 formally defines the problem and establishes the computational complexities of the problem. Section 3 presents the proposed algorithms. We report on the empirical studies in Section 4. Finally, we cover the related work in Section 5 and offer conclusions in Section 6.

## 2. PROBLEM STATEMENT

We define the problem of the keyword-aware optimal route (KOR) query, and show the hardness of the problem.

**Definition 1: Graph.** A graph $\mathcal{G} = (V, E)$ consists of a set of nodes $V$ and a set of edges $E \subseteq V \times V$. Each node $v \in V$ represents a location associated with a set of keywords denoted by $v.\psi$; each edge in $E$ represents a directed route between two locations in $V$, and the edge from $v_i$ to $v_j$ is represented by $(v_i, v_j)$. □

We define $\mathcal{G}$ as a general graph. It can be a road network graph, or a graph extracted from users' historical trajectories. Depending on the source of $\mathcal{G}$, each edge in $\mathcal{G}$ is associated with different types of attributes. For example, if $\mathcal{G}$ is a traffic network, the attributes can be travel duration, travel distance, popularity, and travel cost. To keep our discussion simple, we consider directed graphs only in this paper. However, our discussion can be extended to undirected graphs straightforwardly.

**Definition 2: Route.** A route $R = (v_0, v_1, ..., v_n)$ is a path such that $R$ goes through $v_0$ to $v_n$ sequentially, following the relevant edges in $\mathcal{G}$. □

We define the optimal route based on two attributes on each edge $(v_i, v_j)$: 1) one attribute is used as the *objective value* of this edge, and it is denoted by $o(v_i, v_j)$ (e.g., the popularity), and 2) the other attribute is used as the *budget value* of this edge, which is denoted by $b(v_i, v_j)$ (e.g., the travel time). Note that we can pick up any two attributes to define the optimal route depending on different applications.

**Definition 3: Objective Score and Budget Score.** Given a route $R = \langle v_0, v_1, ..., v_n \rangle$, the objective score of $R$ is defined as the sum of the objective values of all the edges in $R$, i.e.,

$$\mathsf{OS}(R) = \sum_{i=1}^{n} o(v_{i-1}, v_i),$$

and the budget score is defined as the sum of the budget values of all the edges in $R$, i.e.,

$$\mathsf{BS}(R) = \sum_{i=1}^{n} b(v_{i-1}, v_i). \qquad \square$$

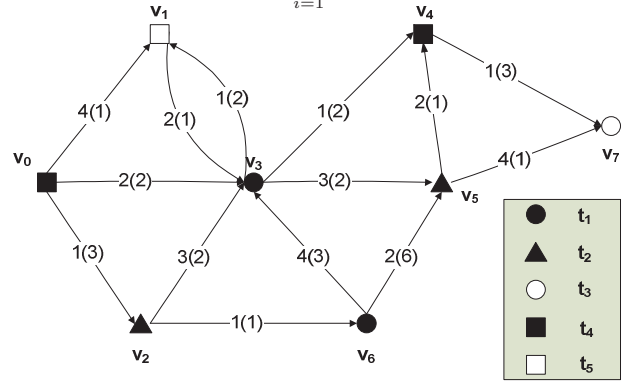

**Figure 1: Example of $\mathcal{G}$**

Figure 1 shows an example of the graph $\mathcal{G}$. We consider only five keywords ($t_1$–$t_5$), and each keyword is represented by a distinct shape. For simplicity, each node contains a single keyword in the example. On each edge, the score inside a bracket is the budget value, and the other number is the objective value. For example, given the route $R = \langle v_0, v_3, v_5, v_7 \rangle$, we have $\mathsf{OS}(R) = 2 + 3 + 4 = 9$ and $\mathsf{BS}(R) = 2 + 2 + 1 = 5$.

Intuitively, a *keyword-aware optimal route* (KOR) query is to find an optimal route from a source to a target in a graph such that the route covers all the query keywords, its budget score satisfies a given constraint, and its objective score is optimized. Formally, we define the KOR query as follows:

**Definition 4: Keyword-aware Optimal Route (KOR) Query.** Given $\mathcal{G}$, the keyword-aware optimal route query $\mathcal{Q} = \langle v_s, v_t, \psi, \Delta \rangle$, where $v_s$ is the source location, $v_t$ is the target location, $\psi$ is a set of keywords, and $\Delta$ specifies the budget limit, aims to find the route $R$ starting from $v_s$ and ending at $v_t$ (i.e., $\langle v_s, \cdots, v_t \rangle$) such that

$$R = \arg\min_R \mathsf{OS}(R)$$
$$\text{subject to} \quad \psi \subseteq \bigcup_{v \in R}(v.\psi)$$
$$\mathsf{BS}(R) \leq \Delta$$

□

In the example graph in Figure 1, given a query $\mathcal{Q} = \langle v_0, v_7, \{t_1, t_2, t_3\}, 8 \rangle$, the optimal route is $R_{opt} = \langle v_0, v_3, v_4, v_7 \rangle$ with objective score $\mathsf{OS}(R_{opt}) = 4$ and budget score $\mathsf{BS}(R_{opt}) = 7$. If we set $\Delta$ to 6, the optimal route becomes $R_{opt} = \langle v_0, v_3, v_5, v_7 \rangle$ with $\mathsf{OS}(R_{opt}) = 9$ and $\mathsf{BS}(R_{opt}) = 5$.

**Theorem 1:** The problem of solving KOR queries is NP-hard.

**Proof Sketch:** This problem can be reduced from the NP-hard weight-constrained shortest path problem (WCSPP) [10]. Given a graph in which each edge has a length and a weight, WCSPP finds a path that has the shortest length with the total weight not exceeding a specified value. The problem of answering KOR queries is a generalization of WCSPP. If each node already covers all the query keywords, the problem of solving KOR becomes equivalent to the WCSPP. □

Obviously, if we disregard the query keyword constraint, the problem of solving KOR becomes WCSPP. In addition, if we remove the budget constraint, the problem becomes similar to the generalized traveling salesman problem (GTSP) [11], which is also NP-hard. In GTSP, the nodes of a graph are clustered into groups, and GTSP finds a path starting and ending at two specified nodes such that it goes through each group exactly once and has the smallest length. In the problem of solving KOR, we can extract the locations whose keywords overlap with $\psi$, and the locations that cover the same keyword form a group. Thus, the problem of solving KOR without the budget constraint is equivalent to the GTSP. Furthermore, if we disregard the objective score, the problem of finding a route that covers all the query keywords and satisfies the budget constraint is still intractable. It is obvious that the simplified problem is also equivalent to GTSP, and thus cannot be solved by polynomial-time algorithms. Many approaches have been proposed for solving GTSP and WCSPP (e.g., [5,7,8,23]. However, they cannot be applied to answer the KOR queries since one more constraint or objective must be satisfied in KOR compared with GSTP and WCSPP.

In the KOR problem, we consider two hard constraints, namely, the keyword coverage and the budget limit, and aim to minimize the objective score. The simplified versions that consider any two aspects are also NP-hard as we analyzed. Hence, it is challenging to find an efficient solution to answering KOR queries. If a route satisfies the two hard constraints, the route is called a *feasible solution* or a *feasible route*.

Furthermore, we can extend the KOR query to the *keyword-aware top-k route* (K$k$R) query. Instead of finding the optimal route defined in KOR, the K$k$R query is to return $k$ routes starting and ending at the given locations such that they have the smallest objective scores, cover the query keywords, and satisfy the given budget constraint.

## 3. ALGORITHMS

We present the pre-processing method in Section 3.1, the proposed approximation algorithm OSScaling with provable approximation bound in Section 3.2, the more efficient approximation algorithm BucketBound also with performance guarantee in Section 3.3, and the greedy algorithm Greedy in Section 3.4.

### 3.1 Pre-processing

We introduce the pre-processing method. We utilize the pre-processing results to accelerate the algorithms to be proposed.

We use the Floyd-Warshall algorithm [9], which is a well-known algorithm for finding all pairs shortest path, to find the following two paths for each pair of nodes $(v_i, v_j)$:

- $\tau_{i,j}$: the path with the smallest objective score. The objective score of this path is denoted by $\mathsf{OS}(\tau_{i,j})$ and the budget score is denoted by $\mathsf{BS}(\tau_{i,j})$.
- $\sigma_{i,j}$: the path with the smallest budget score. The objective score of $\sigma_{i,j}$ is denoted by $\mathsf{OS}(\sigma_{i,j})$ and the budget score is denoted by $\mathsf{BS}(\sigma_{i,j})$.

For example, after the pre-processing, for the pair of node $(v_0, v_7)$ in Figure 1, we have $\tau_{0,7} = \langle v_0, v_3, v_4, v_7 \rangle$ with $\mathsf{OS}(\tau_{0,7}) = 4$ and $\mathsf{BS}(\tau_{0,7}) = 7$ and $\sigma_{0,7} = \langle v_0, v_3, v_5, v_7 \rangle$ with $\mathsf{OS}(\sigma_{0,7}) = 9$ and $\mathsf{BS}(\sigma_{0,7}) = 5$.

Only the objective and budget scores of $\tau_{i,j}$ and $\sigma_{i,j}$ are used in the proposed algorithms, while the two paths themselves are not. The space cost is $O(|V|^2)$, where $|V|$ represents the number of nodes in the graph. In general, the number of points of interests $|V|$ within a city is not large [15, 19].

We use an inverted file to organize the word information of nodes. An inverted file index has two main components: 1) A vocabulary of all distinct words appearing in the descriptions of nodes (locations), and 2) A posting list for each word $t$ that is a sequence of identifiers of the nodes whose descriptions contain $t$. We use $B^+$-tree for the inverted file index, which is disk resident.

### 3.2 Approximation Algorithm OSScaling

A brute-force approach to solving KOR is to do an exhaustive search: We enumerate all candidate paths from the source node. We can use a queue to store the partial paths. In each step, we select one partial path from the queue. Then it is extended to generate more candidate partial paths and those paths whose budget scores are smaller than the specified limit are enqueued. When a path is extended to the target node, we check whether it covers all the query keywords and satisfies the budget constraint. We record all the feasible routes, and after all the candidate routes from the source node to the target node have been checked, we select the best one of all the feasible routes as the answer to the query.

However, the exhaustive search is computationally prohibitive. Given a query with a specified budget limit $\Delta$, we know that the number of edges in a route exploited in the search is at most $\lfloor \frac{\Delta}{b_{min}} \rfloor$, where $b_{min}$ is the smallest budget value of all edges in $\mathcal{G}$. Thus, the complexity of an exhaustive search is $O(d^{\lfloor \frac{\Delta}{b_{min}} \rfloor})$, where $d$ is the maximum outdegree in $\mathcal{G}$ (notice that enumerating all the simple paths is not enough for answering KOR queries). To avoid the expensive exhaustive search, we devise a novel approximation algorithm OSScaling. It is challenging to develop such an algorithm.

The main problem of the brute-force approach is that too many partial paths need to be stored on each node. In order to reduce the cost of enumerating the partial paths, in OSScaling, we scale the objective values of edges in $\mathcal{G}$ into integers utilizing a parameter $\epsilon$. The scaling enables us to bound the number of partial paths explored, and further to design a novel algorithm that runs polynomially in the budget constraint $\Delta$, $\frac{1}{\epsilon}$, the number of nodes and edges in $\mathcal{G}$, and is exponential in the number of query keywords (which is typically small). Furthermore, the objective score scaling guarantees that the algorithm always returns a route whose objective score is no more than $\frac{1}{1-\epsilon}$ times of that of the optimal route, if there exists one. This is inspired by the FPTAS (fully polynomial-time approximation scheme) for solving the well-known knapsack problem [24]. Note that the problem of answering KOR queries is different from the NP-hard problem knapsack and its solutions cannot be used.

We define a scaling factor $\theta = \frac{\epsilon o_{min} b_{min}}{\Delta}$, where $o_{min}$ and $b_{min}$ represent the smallest objective value and the smallest budget value of all edges in $\mathcal{G}$, respectively, and $\epsilon$ is a parameter in the range $(0, 1)$. Next, for each edge $(v_i, v_j)$, we scale its objective value $o(v_i, v_j)$ to $\hat{o}(v_i, v_j) = \lfloor \frac{o(v_i,v_j)}{\theta} \rfloor$. We call the graph with scaled objective values as the **scaled graph**, denoted by $\mathcal{G}_S$. Given a route $R = \langle v_0, v_1, ..., v_n \rangle$ in $\mathcal{G}_S$, we denote its scaled objective score by $\hat{OS}(R) = \sum_{i=1}^{n} \hat{o}(v_{i-1}, v_i)$.

On the scaled graph, we still extend from the source node to create new partial paths until we reach the target node. However, if a partial path has both smaller scaled objective score and budget score than another one on the same node, the OSScaling algorithm ignores it. Before detailing the algorithm, we introduce the following important definitions.

**Definition 5: Node Label.** For each node $v_i$, we maintain a list of labels, in which each label corresponds to a path $P_i^k$ from the source node $v_s$ to node $v_i$. The label is denoted by $L_i^k$ and is in format of $(\lambda, \hat{OS}, OS, BS)$, where $L_i^k.\lambda$ is the keywords covered by $P_i^k$, $L_i^k.\hat{OS}$, $L_i^k.OS$, and $L_i^k.BS$ represent the scaled objective score, the original objective score, and the budget score of $P_i^k$, respectively. □

**Example 1:** In the example graph shown in Figure 1, assuming $\Delta = 10$ and $\epsilon = 0.5$, we can compute the value for $\theta$: $\theta = \frac{0.5*b_{min}*o_{min}}{10} = \frac{1}{20}$. Therefore, the objective value of each edge is scaled to 20 times of its original value. Given the two paths from $v_0$ to $v_4$, i.e., $R_1 = \langle v_0, v_2, v_3, v_4 \rangle$ and $R_2 = \langle v_0, v_2, v_6, v_5, v_4 \rangle$. The label of $R_1$ is $L_4^0 = (\langle t_1, t_2, t_4 \rangle, 100, 5, 7)$ and the label of $R_2$ is $L_4^1 = (\langle t_1, t_2, t_4 \rangle, 120, 6, 11)$. □

Each partial route is represented by a node label. At each node, we maintain a list of labels, each of which stores the information of a corresponding partial route from the source node to this node, including the query keywords already covered, the scaled objective score, the original objective score, and the budget score of the partial route. Many paths between two nodes may exist, and thus each node may be associated with a large number of labels. However, most of the labels are not necessary for answering KOR. Considering Example 1, at node $v_4$, the label $L_4^1$ could be ignored since $L_4^0$ has both smaller objective and budget scores. This is because that in the route extended from $L_4^1$, we can always replace the partial route corresponding to $L_4^1$ with that corresponding to label $L_4^0$. We say that $L_4^0$ **dominates** $L_4^1$:

**Definition 6: Label Domination.** Let $L_i^k$ and $L_i^l$ be two labels corresponding to two different paths from the source node $v_s$ to node $v_i$. We say $L_i^k$ **dominates** $L_i^l$ iff $L_i^k.\lambda \supseteq L_i^l.\lambda$, $L_i^k.\hat{OS} \leq L_i^l.\hat{OS}$, and $L_i^k.BS \leq L_i^l.BS$. □

Notice that in OSScaling we determine if a label dominates another one with regard to the scaled objective score instead of the original objective score. Therefore, it is likely that the label dominated has smaller original objective score, and hence the optimal route may be missed in this algorithm. This is the reason that OSScaling can only return approximate results. However, by doing so, the maximum number of labels on a node is bounded, which further bounds the complexity of OSScaling. We have the following lemma:

**Lemma 1:** On a node there are at most $2^m \lfloor \frac{\Delta}{b_{min}} \rfloor \lfloor \frac{o_{max}\Delta}{\epsilon o_{min}b_{min}} \rfloor$ labels, where $m$ is the number of query keywords, $\epsilon$ is the scaling parameter, $b_{min}$, $o_{max}$, and $o_{min}$ represent the smallest budget value, the largest objective value, and the smallest objective value of all edges in $\mathcal{G}$, respectively.

**Proof Sketch:** First, given $m$ query keywords, there are at most $2^m$ keywords subset. Second, given the budget limit $\Delta$, the number of edges in a route checked by our algorithm does not exceed $\lfloor \frac{\Delta}{b_{min}} \rfloor$. Hence, the objective score of a route in $\mathcal{G}_S$ is bounded by $\lfloor \frac{\Delta}{b_{min}} \rfloor \hat{o}_{max} = \lfloor \frac{\Delta}{b_{min}} \rfloor \lfloor \frac{o_{max}}{\theta} \rfloor = \lfloor \frac{\Delta}{b_{min}} \rfloor \lfloor \frac{o_{max}\Delta}{\epsilon o_{min}b_{min}} \rfloor$. In conclusion, we only need to store at most $2^m \lfloor \frac{\Delta}{b_{min}} \rfloor \lfloor \frac{o_{max}\Delta}{\epsilon o_{min}b_{min}} \rfloor$ labels, because all the rest can be dominated by them. □

Note that Lemma 1 gives an upper bound of the label number at a node. In practice, the number of labels maintained at a node is usually much smaller than this upper bound. We denote this upper bound by $L_{max}$.

Next, we introduce how to do the route extension using labels. This step is called *label treatment*:

**Definition 7: Label Treatment.** Given a label $L_i^k$ at node $v_i$, for each outgoing neighbor $v_j$ of node $v_i$ in $\mathcal{G}$, we create a new label for $v_j$: $L_j^t = (L_i^k.\lambda \bigcup v_j.\psi, L_i^k.\hat{OS} + \hat{o}(v_i, v_j), L_i^k.OS + o(v_i, v_j), L_i^k.BS + b(v_i, v_j))$. □

The label treatment step extends a partial route at node $v_i$ forward to all the outgoing neighbor nodes of $v_i$, and thus more longer partial routes are generated. Note that the label treatment step is applied together with label domination checking.

Another important definition is how we compare the order of two labels:

**Definition 8: Label Order.** Let $L_i^k$ and $L_j^t$ be two labels corresponding to two paths from source node $v_s$ to node $v_i$ and $v_j$ ($v_i$ and $v_j$ can be either the same or different nodes), respectively. We say $L_i^k$ has a lower order than $L_j^t$, denoted by $L_i^k \prec L_j^t$, iff $|L_i^k.\lambda| > |L_j^t.\lambda|$ or $(|L_i^k.\lambda| = |L_j^t.\lambda|$ and $L_i^k.\hat{OS} < L_j^t.\hat{OS})$ or $(|L_i^k.\lambda| = |L_j^t.\lambda|, L_i^k.\hat{OS} = L_j^t.\hat{OS}$, and $L_i^k.BS < L_j^t.BS)$; otherwise, breaking the tie by alphabetical order of $v_i$ and $v_j$. □

In Example 1, we say that $L_4^0 \prec L_4^1$, because they contain the same number of query keywords, and $L_4^0$ has smaller objective and budget scores. This definition decides which partial route is selected for extension in each step.

Now we are ready to present our algorithms. The basic idea is to keep creating new partial routes from the best one among all existing partial routes. From the viewpoint of node labels, we first create a label at the source node, and then we keep generating new labels that cannot be dominated by existing ones. We always select the one with the smallest order according to Definition 8 to generate new labels. If newly generated labels cannot be dominated by existing labels, they are used to detect and delete the labels dominated by them. We repeat this procedure until all the labels on the target node are generated, and finally the label with the best objective score satisfying the budget limit at the target node is returned. Note that this is not an exhaustive search algorithm and we will analyze the complexity after presenting the algorithm.

The pseudocode is presented in Algorithm 1. We use a min-priority queue $Q$ to organize the labels, which are enqueued into $Q$ according to their orders defined in Definition 8. We use variable $U$ to keep track of the upper bound of the objective score, and use $LL$ to store the last label of the current best route. We initialize $U$ as $\infty$, and set $LL$ as $NULL$. We create a label at the starting node $v_s$ and enqueue it into $Q$ (lines 2–4).

We keep dequeuing labels from $Q$ until $Q$ becomes empty (lines 5–20). We terminate the algorithm when $Q$ is empty or when all the labels in $Q$ has objective scores larger than $U$. In each while-loop, we first dequeue a label $L_i^k$ with the minimum label order from $Q$ (line 6). If the objective score of $L_i^k$ plus the best objective score $\text{OS}(\tau_{i,t})$ from $v_i$ to the target node $v_t$ is larger than the current upper bound $U$, then the label definitely cannot contribute to the final result (line 7). Next, for each outgoing neighbor $v_j$ of $v_i$, we create a new label $L_j^l$ for it according to Definition 7 (line 9). If $L_j^l$ can be dominated by other labels on the node $v_j$ or if it cannot generate a

**Algorithm 1:** OSScaling Algorithm

1  Initialize a min-priority queue $Q$;
2  $U \leftarrow \infty$; $LL \leftarrow NULL$;
3  At node $v_s$, create a label: $L_s^0 \leftarrow (v_s.\psi, 0, 0, 0)$;
4  $Q$.enqueue($L_s^0$);
5  **while** $Q$ is not empty **do**
6     $L_i^k \leftarrow Q$.dequeue();
7     **if** $L_i^k.OS + \mathsf{OS}(\tau_{i,t}) > U$ **then continue**;
8     **for** each edge $(v_i, v_j)$ **do**
9       Create a label $L_j^l$ for $v_j$: $L_j^l \leftarrow (L_i^k.\lambda \bigcup v_j.\psi, L_i^k.\hat{OS} + \hat{o}(v_i, v_j), L_i^k.OS + o(v_i, v_j), L_i^k.BS + b(v_i, v_j))$;
10      **if** $L_j^l$ is not dominated by other labels on $v_j$ **and** $L_j^l.BS + \mathsf{BS}(\sigma_{j,t}) < \Delta$ **and** $L_j^l.OS + \mathsf{OS}(\tau_{j,t}) < U$ **then**
11        **if** $L_j^l$ does not cover all the query keywords **then**
12          $Q$.enqueue($L_j^l$);
13          **for** each label $L$ on $v_j$ **do**
14            **if** $L$ is dominated by $L_j^l$ **then**
15              remove $L$ from $Q$;
16       **else**
17          **if** $L_j^l.BS + \mathsf{BS}(\tau_{j,t}) < \Delta$ **then**
18             $U \leftarrow L_j^l.OS + \mathsf{OS}(\tau_{j,t})$;
19             $LL \leftarrow L_j^l$;
20          **else** $Q$.enqueue($L_j^l$);
21 **if** $U$ is $\infty$ **then** return "No feasible route exits";
22 **else** Obtain the route utilizing $LL$ and return it;

feasible route (first, the budget score of $L_j^l$ plus $\mathsf{BS}(\sigma_{j,t})$, the best budget score to $v_t$, is larger than the budget constraint $\Delta$; second, the objective score of $L_j^l$ plus $\mathsf{OS}(\tau_{j,t})$, the best objective score to $v_t$, is larger than the current upper bound $U$), we ignore the new label (line 10); Otherwise, if it does not cover all the query keywords, we enqueue it into $Q$ and use it to detect and delete the labels that are dominated by it on $v_j$ (lines 11–15).

When we find that the current label $L_j^l$ already covers all the query keywords, a feasible solution is found and we update the upper bound $U$ (lines 16–20). First, if the budget score of $L_j^l$ plus the budget score of $\tau_{j,t}$ (the path with the best objective score from $v_j$ to $v_t$) is smaller than $U$, we update the upper bound $U$, and the last label is also updated (lines 18–19); otherwise, we enqueue this label into $Q$ for later processing. Finally, if $U$ is never updated, we know that there exists no feasible route for the given query; otherwise, we can construct the route using the label $LL$ (lines 21–22).

The following example illustrates how this algorithm works.

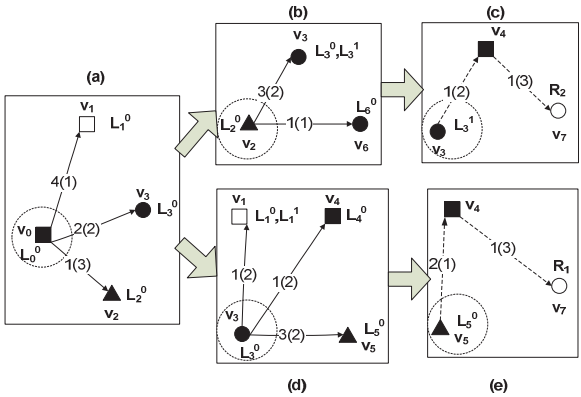

**Figure 2:** Steps of Example 2

**Example 2:** Consider the example graph in Figure 1, the query $\mathcal{Q} = \langle v_0, v_7, \{t_1, t_2\}, 10 \rangle$, and $\epsilon$ is set as 0.5. The steps of the algorithm are shown in Figure 2 and the contents of the labels generated are in Table 1.

Initially, we create a label $L_0^0=(\emptyset, 0, 0, 0)$ at node $v_0$ and enqueue it into $Q$. After we dequeue it from $Q$, as shown in step (a), we generate the following three labels on all the outgoing neighbors of $v_0$: $L_1^0$, $L_2^0$, and $L_3^0$. The three labels are also enqueued into $Q$.

In the next loop, $L_2^0$ is selected because $L_2^0 \prec L_3^0 \prec L_1^0$. As shown in Step (b), we generate another two labels $L_3^1$ and $L_6^0$. Note that the best budget score from $v_6$ to $v_7$ is 7 ($\mathsf{BS}(\sigma_{6,7})$=7), and thus $L_6^0$ can be ignored since $L_6^0.BS + \mathsf{BS}(\sigma_{6,7})$ (=11)> $\Delta$. According to the pre-processing results, $\mathsf{OS}(\tau_{3,7})$=2 and $\mathsf{BS}(\tau_{3,7})$ =5. Therefore, in step (c), we get a feasible route $R_1 = \langle v_0, v_2, v_3, v_4, v_7 \rangle$ with $\mathsf{OS}(R_1)$ =6 and $\mathsf{BS}(R_1)$=10. The upper bound $U$ is updated as $\mathsf{OS}(R_1)$, i.e., $U$=6.

Next, $L_3^0$ on node $v_3$ is selected. As shown in Step (d), we generate another three labels and enqueue them into $Q$: $L_1^1$, $L_4^0$, and $L_5^0$. Now label $L_5^0$ already covers all the query keywords on $v_5$. According to the pre-processing results, from $v_5$ to $v_7$, the best objective score is 3 ($\mathsf{OS}(\tau_{5,7})$=3) and the budget score of this path is 4. Utilizing the pre-processing results, as shown in step (e), we can obtain another feasible solution $R_2 = \langle v_0, v_3, v_5, v_4, v_7 \rangle$ with $\mathsf{OS}(R_2)$=8 and $\mathsf{BS}(R_2)$(=8) < $\Delta$ (Note that suppose $\Delta$=7 in $\mathcal{Q}$, $R_2$ will not be a feasible result. Instead, we enqueue the label $L_5^0$ into $Q$, and in the next loop, we include the edge $(v_5, v_7)$ and get a feasible route $\langle v_0, v_3, v_5, v_7 \rangle$).

The rest labels are treated similarly, and the best route is $R_1$. □

| | $L_0^0$ | $L_1^0$ | $L_1^1$ | $L_2^0$ | $L_3^0$ | $L_3^1$ | $L_4^0$ | $L_5^0$ | $L_6^0$ |
|---|---|---|---|---|---|---|---|---|---|
| $\lambda$ | $\emptyset$ | $\emptyset$ | $t_1$ | $t_2$ | $t_1$ | $t_1,t_2$ | $t_1$ | $t_1,t_2$ | $t_1,t_2$ |
| $\hat{OS}$ | 0 | 80 | 60 | 20 | 40 | 80 | 60 | 100 | 40 |
| $OS$ | 0 | 4 | 3 | 1 | 2 | 4 | 3 | 5 | 2 |
| $BS$ | 0 | 1 | 4 | 3 | 2 | 5 | 4 | 4 | 4 |

**Table 1:** Labels contents

**Complexity:** In each loop of OSScaling, we dequeue one label from $Q$. Thus, in the worst case we need $|V|L_{max}$ loops according to Lemma 1. Within one loop, 1) we generate new labels on a node and check the domination on its outgoing neighbors, taking $O(|E|L_{max})$ time by aggregate analysis; 2) we dequeue one label and the complexity is $O(lgL_{max})$. Hence, we can conclude that the worst time complexity is $O(|V|L_{max}lgL_{max} + |E|L_{max}))$. In practice, the number of loop is much smaller than the worst case and the number of keywords of a query is quite small. Therefore, the algorithm OSScaling is able to return the result efficiently.

By scaling the objective values of edges in $\mathcal{G}$, the algorithm OSScaling is able to guarantee an approximation bound.

**Approximation Bound:** We denote the route found by OSScaling as $R_{OS}$, and the feasible route with the smallest scaled objective score in $\mathcal{G}_S$ as $R_{\mathcal{G}_S}$. We have the following lemma:

**Lemma 2:** $\mathsf{OS}(R_{\mathcal{G}_S}) \geq \mathsf{OS}(R_{OS})$.

**Proof Sketch:** In Algorithm 1, if we use the partial route with the smallest scaled objective score to update the upper bound at node $v_j$ (line 18), the algorithm returns $R_{\mathcal{G}_S}$. We denote the objective score of a route from $v_p$ to $v_q$ as $O_{p,q}$, and we know $O_{s,j}(R_{\mathcal{G}_S}) = O_{s,j}(R_{OS})$. According to the algorithm, $O_{j,t}(R_{\mathcal{G}_S}) \geq \tau_{j,t} = O_{j,t}(R_{OS})$, and thus $\mathsf{OS}(R_{\mathcal{G}_S}) = O_{s,j}(R_{\mathcal{G}_S}) + O_{j,t}(R_{\mathcal{G}_S}) \geq \emptyset_{s,j}(R_{OS}) + O_{j,t}(R_{OS}) = \mathsf{OS}(R_{OS})$. □

We denote the optimal route as $R_{opt}$. We have:

**Theorem 2:** $\mathsf{OS}(R_{opt}) \geq (1-\epsilon)\mathsf{OS}(R_{OS})$.

**Proof Sketch:** From $\hat{o} = \lfloor \frac{o}{\theta} \rfloor$, we know that $o - \theta \leq \theta\hat{o} \leq o$. Therefore, $\mathsf{OS}(R_{opt}) = \sum_{e \in R_{opt}} o_e \geq \sum_{e \in R_{opt}} \theta\hat{o}_e$., then:

$$\mathsf{OS}(R_{opt}) \geq \sum_{e \in R_{opt}} \theta\hat{o}_e \geq \sum_{e' \in R_{\mathcal{G}_S}} \theta\hat{o}_{e'} \geq \sum_{e' \in R_{\mathcal{G}_S}} (o_{e'} - \theta)$$

$$\geq \sum_{e' \in R_{\mathcal{G}_S}} o_{e'} - \lfloor \frac{\Delta}{b_{min}} \rfloor \theta \geq \sum_{e' \in R_{\mathcal{G}_S}} o_{e'} - \epsilon o_{min}$$

Because $\sum_{e' \in R_{\mathcal{G}_S}} o_{e'} \geq o_{min}$, we can conclude that $\mathsf{OS}(R_{opt}) \geq (1-\epsilon) \sum_{e' \in R_{\mathcal{G}_S}} o_{e'} = (1-\epsilon)\mathsf{OS}(R_{\mathcal{G}_S}) \geq (1-\epsilon)\mathsf{OS}(R_{OS})$ (according to Lemma 2). □

We can see that the parameter $\epsilon$ affects not only the running time of this algorithm but also the accuracy. There is a tradeoff between the efficiency and accuracy when selecting a value for $\epsilon$. With a larger value of $\epsilon$, OSScaling runs faster but the accuracy would drop; on the contrary, with a smaller value for $\epsilon$ we can obtain better routes but that needs longer query time.

**Optimization:** We design the following optimization strategies to further improve Algorithm 1.

**Optimization Strategy 1:** When processing a label $L_i^k$ at node $v_i$, in addition to the labels generated by following the outgoing edges of $v_i$ in the graph, we also generate a label on a node $v_j$ such that $\mathsf{BS}(\sigma_{i,j})$ has the smallest value among all the nodes containing a uncovered query keyword and $L_i^k.BS + \mathsf{BS}(\sigma_{i,j}) + \mathsf{BS}(\sigma_{j,t}) \leq \Delta$. The motivation of this strategy is to find a feasible solution as early as possible, and then it is used to update the upper bound and further to prune more labels.

**Optimization Strategy 2:** When the query contains some very infrequent words, we can utilize the nodes that contain them to find the result more efficiently. In Algorithm 1, when we decide if a label $L_i^k$ can be deleted, two specific conditions are checked: 1) if $L_i^k.OS + \mathsf{OS}(\tau_{i,t})$ is smaller than $U$; 2) if $L_i^k.BS + \mathsf{BS}(\sigma_{i,t})$ is smaller than $\Delta$. We utilize the scores of the two pre-processed routes from $v_i$ to the target node $v_t$. But if the path from $v_i$ to the nodes containing the infrequent words have large objective or budget scores, we will waste a lot of time on extending the route from $v_i$. The reason is that, although the label $L_i^k$ cannot be pruned by the two conditions, it cannot generate useful labels, and this is not known until we reach the nodes containing the infrequent words. We first obtain all the nodes containing the least infrequent word (which must be below a frequency threshold, such as appearing in less than 1% nodes) utilizing the inverted file; after we generate a label $L_i^k$, if it does not cover the least infrequent word, for each node $l$, we check two conditions: 1) $L_i^k.OS + \mathsf{OS}(\tau_{i,l}) + \mathsf{OS}(\tau_{l,t}) > U$; 2) $L_i^k.BS + \mathsf{BS}(\sigma_{i,l}) + \mathsf{BS}(\sigma_{l,t}) > \Delta$. If on each node containing infrequent words at least one condition is satisfied, this label can be discarded.

### 3.3 Approximation Algorithm BucketBound

In the algorithm OSScaling, after we find a feasible solution, we still have to keep searching for a better route until all the feasible routes are checked. We propose a more efficient approximate method denoted by BucketBound with provable approximation bounds which is also based on scaling the objective scores into integers.

Before describing the proposed algorithm, we introduce the following lemma which lays a foundation of this algorithm.

**Lemma 3:** Given a label $L_i^k$ at node $v_i$, the best possible objective score of the feasible routes that could be extended from the partial path represented by $L_i^k$ is $L_i^k.OS + \mathsf{OS}(\tau_{i,t})$. We denote the score by $\mathsf{LOW}(L_i^k)$.

**Proof Sketch:** If $\tau_{i,t}$ and $L_i^k$ cover all query keywords collectively, they constitute the best route extending from $L_i^k$ and its objective score is equal to $L_i^k.OS + \mathsf{OS}(\tau_{i,t})$. Otherwise, another route from $v_i$ to $v_t$ covering more keywords must be selected to construct a feasible route. This route has larger objective score than that of $\tau_{i,t}$, which results in a larger objective score of the final route. □

In this algorithm, we divide the traversed partial routes into different "buckets" according to their best possible objective scores. We define the buckets as follows:

**Definition 9: Label Buckets.** The label buckets organize labels. Each bucket is associated with an order number and corresponds to an objective score interval—the $r$th bucket $B_r$ corresponds to the following interval: $[\beta^r \mathsf{OS}(\tau_{s,t}), \beta^{r+1}\mathsf{OS}(\tau_{s,t}))$, where $\mathsf{OS}(\tau_{s,t})$ is the best objective score from $v_s$ to $v_t$ and $\beta$ is a specified parameter. A label $L_i^k$ is in the bucket $B_r$ if:

$$\beta^r \mathsf{OS}(\tau_{s,t}) \leq \mathsf{LOW}(L_i^k) < \beta^{r+1}\mathsf{OS}(\tau_{s,t})$$

□

With this important definition, we proceed to present the approximation algorithm BucketBound. We denote the route found by OSScaling as $R_{OS}$. The basic idea is as follows: We keep selecting labels (partial routes) from the buckets. When selecting a label, we always choose the non-empty bucket with the smallest order number, and then select a label with the lowest label order from it. After a label $L_i^k$ is generated, we compute the score $\mathsf{LOW}(L_i^k)$ and we place this label to the corresponding bucket according to Definition 9. Utilizing the label buckets enables us to find a novel way to detect if a feasible route found is in the same bucket as $R_{OS}$. If we find such a route during the above procedure, we return it as the result. We denote the route found by BucketBound as $R_{BB}$.

We proceed to explain how to determine if the bucket where we find a feasible route contains $R_{OS}$.

This algorithm follows the basic label generation and selection approach in OSScaling. However, the strategies of generating and selecting labels are different. With such changed label generation and selection strategies, we have the following lemma:

**Lemma 4:** If all the buckets $B_i(i = 0, ..., r)$ are empty and no feasible solution is found yet, the objective score of $R_{OS}$ satisfies: $\mathsf{OS}(R_{OS}) \geq \beta^{r+1}\mathsf{OS}(\tau_{s,t})$.

**Proof Sketch:** Since any bucket $B_i(i \leq r)$ is empty, we know the label corresponding to $R_{OS}$ must be selected from the subsequent buckets. Therefore, $\mathsf{LOW}(L_j^l) > \beta^{r+1}\mathsf{OS}(\tau_{s,t})$. According to Lemma 3, we know $\mathsf{OS}(R_{OS}) \geq \mathsf{LOW}(L_j^l) \geq \beta^{r+1}\mathsf{OS}(\tau_{s,t})$. □

Based on Lemma 4, we have Lemma 5. When the condition in Lemma 5 is satisfied, a feasible route and $R_{OS}$ fall into the same bucket, and the algorithm terminates.

**Lemma 5:** When a feasible route $R_{BB}$ is found in the bucket $B_{r+1}$ and all the buckets $B_0, B_1, ..., B_r$ are empty, the route $R_{OS}$ found by OSScaling is also contained in $B_{r+1}$.

**Proof Sketch:** Because any bucket $B_i(i \leq r)$ is empty, according to Lemma 4, $\mathsf{OS}(R_{OS}) \geq \beta^{r+1}\mathsf{OS}(\tau_{s,t})$. Since $\mathsf{OS}(R_{OS}) \leq \mathsf{OS}(R_{BB})$ ($R_{BB}$ is one feasible solution found in OSScaling), we know $\beta^{r+1}\mathsf{OS}(\tau_{s,t}) \leq \mathsf{OS}(R_{OS}) \leq \mathsf{OS}(R_{BB}) < \beta^{r+2}\mathsf{OS}(\tau_{s,t})$. According to Definition 9, $R_{OS}$ also falls in $B_{r+1}$. □

Figure 3 illustrates the basic process of the proposed approximation algorithm BucketBound. As shown in the figure, we first select the label $L_i^k$ from the bucket $B_0$, and after the label treatment the new label is put into the bucket $B_3$. Since $B_0$ becomes empty now, we proceed to select labels from $B_1$. If $B_0$, $B_1$, and $B_2$ all become empty, according to Lemma 4 we can know $\mathsf{OS}(R_{OS}) \geq \beta^3 \mathsf{OS}(\tau_{s,t})$. If now we find a feasible route $R_{BB}$ in the bucket $B_3$, according to Lemma 5, it is assured that $R_{OS}$ also falls into $B_3$, and we return $R_{BB}$ as the result.

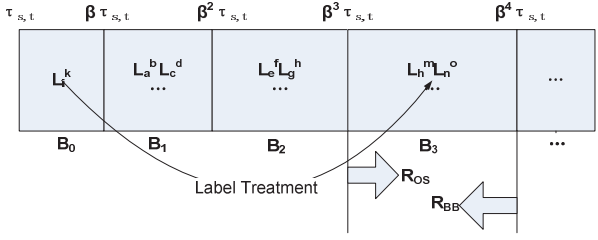

**Figure 3: Process of Algorithm 2**

Unlike Algorithm 1, the approximation algorithm terminates immediately when Lemma 5 is satisfied, which means a feasible solution is found. Note that the feasible solution may be different from the first feasible solution found by Algorithm 1. This algorithm is also capable of determining if a feasible route exists. If all buckets are empty during the label selection step and no feasible route found yet, there exists no result for KOR. This is because that when all buckets are empty, all the labels generated do not satisfy the budget constraint, which means that all the partial routes generated from the source node exceed the budget limit $\Delta$.

**Algorithm 2: BucketBound Algorithm**

1 Initialize a min-priority queue $B_0$;
2 $LL \leftarrow NULL$; $Found \leftarrow$ false;
3 At node $v_s$, create label $L_s^0 \leftarrow (v_s.\psi, 0, 0, 0)$;
4 $B_0$.enqueue($L_s^0$);
5 **while** *Found is false* **do**
6    $B_r \leftarrow$ the queue of the first non-empty bucket;
7    **if** *All queues are empty* **then return** "No feasible route exist";
8    $L_i^k \leftarrow B_r$.dequeue();
9    **for** *each edge $(v_i, v_j)$* **do**
10      Create a new label $L_j^l$ for $v_j$:
$L_j^l \leftarrow (L_i^k.\lambda \bigcup v_j.\psi, L_i^k.\hat{OS} + \hat{o}(v_i, v_j), L_i^k.OS + o(v_i, v_j), L_i^k.BS + b(v_i, v_j))$;
11      **if** $L_j^l$ *is not dominated by other labels on $v_j$ and* $L_j^l.BS + \mathsf{BS}(\sigma_{j,t}) < \Delta$ **then**
12        Find $B_s$ that $L_j^l$ falls into;
13        **if** $B_s$ *does not exist* **then**
14          Initialize a priority queue $B_s$;
15        $B_s$.enqueue($L_j^l$);
16        **for** *each label $L$ on $v_j$* **do**
17          **if** $L$ *is dominated by* $L_j^l$ **then**
18            remove $L$ from the corresponding queue;
19        **if** $L_j^l$ *covers all the query keywords* **then**
20          **if** $B_r$ *and* $B_s$ *are the same queue* **then**
21            **if** $L_j^l.BS + \mathsf{BS}(\tau_{j,t}) \leq \Delta$ **then**
22              $Found \leftarrow$ true;// Lemma 5
23              $LL \leftarrow L_j^l$;
24 Obtain the route utilizing $LL$ and return the route;

The algorithm is detailed in Algorithm 2. It uses a min-priority queue for each bucket to organize the labels in the bucket. We initialize the first min-priority queue $B_0$ (corresponding to the first bucket with boundary $[\mathsf{OS}(\tau_{s,t}), \beta \mathsf{OS}(\tau_{s,t}))$); $U$ and $LL$ are initialized as in Algorithm 1. We initialize the flag $Found$ as false, which records if a feasible route is found. We create a label at the source node $v_s$ and enqueue it into $B_0$ (lines 1–4). The algorithm terminates when the flag $Found$ is true. We keep dequeuing labels from $B_r$ which represents the non-empty bucket with the smallest order number until we find a solution or no result exists(lines 5–23). If all queues become empty, it is assured that no feasible route exists (line 7). After we select a label $L_i^k$ on node $v_i$, for each outgoing neighbor $v_j$ of $v_i$, we create a new label for it (line 10). When a new label $L_j^l$ is generated, we check: 1) if it can be dominated by other labels on $v_j$; 2) if it cannot generate results definitely. If so, we ignore it (line 11); Otherwise, we use it to delete labels on $v_j$ that can be dominated by it, and we enqueue this label to the corresponding bucket according to its best possible objective score (lines 12–18). When $L_j^l$ already covers all the query keywords and also falls into $B_r$, we still need to test if the path corresponding to $\mathsf{LOW}(L_j^l)$ satisfies the budget constraint. If so, we find a solution and exit the loop according to Lemma 5 (lines 19–23).

**Theorem 3:** Algorithm 2 offers the approximation ratio $\frac{\beta}{1-\epsilon}$. **Proof Sketch:** Assume that the solution $R_{BB}$ is found in $B_k$. According to Lemma 5, the route found by OSScaling $R_{OS}$ is also contained in $B_k$. Thus, we have $\mathsf{OS}(R_{OS}) \geq \beta^k \mathsf{OS}(\tau_{st})$ and $\mathsf{OS}(R_{BB}) < \beta^{k+1}\mathsf{OS}(\tau_{st})$. According to Theorem 2, we can get: $\frac{\mathsf{OS}(R_{BB})}{\mathsf{OS}(R_{opt})} = \frac{\mathsf{OS}(R_{BB})}{\mathsf{OS}(R_{OS})} \frac{\mathsf{OS}(R_{OS})}{\mathsf{OS}(R_{opt})} < \frac{\beta^{k+1}\mathsf{OS}(\tau_{st})}{\beta^k \mathsf{OS}(\tau_{st})(1-\epsilon)} = \frac{\beta}{1-\epsilon}$ □ Although BucketBound has the same worst case complexity as Algorithm 1, it processes much fewer labels and is more efficient in practice. Note that the two optimization strategies in OSScaling are still applicable in BucketBound.

### 3.4 Greedy Algorithm

We propose an approximation algorithm using the greedy approach to solve KOR. It has no performance guarantee.

There are three constraints in the KOR problem: a) a set of keywords must be covered; b) the objective score must be minimized; c) the budget limit $\Delta$ must be satisfied. As discussed in Section 2, by considering only two of them, the problem is still NP-hard. Therefore, a greedy approach normally cannot grantee that two constraints are satisfied. Since the keyword and budget constraints are hard constraint, we design a greedy algorithm such that it is able to find a route either covering all the query keywords or satisfying the budget constraint, while minimizing the objective score greedily.

The idea is that we start from the source node, and keep selecting the next best node according to a certain strategy until we finally reach the target node. The strategy of selecting the next node affects the results significantly. We design a greedy strategy that takes into account all the three constraints simultaneously to find the best next node: a) the node contains uncovered query keywords; and b) the best route that can be generated after including this node into the current partial route is expected to have a small objective score and fulfill the budget constraint. We use a parameter $\alpha$ to balance the importance of the objective and budget scores when selecting a node: at node $v_i$, when we extend the current partial route $R_i$ ending at $v_i$, we select the node $v_j$ that minimizes the following score:

$$\mathsf{score}(v_j, R_i) = \alpha(R_i.OS + \mathsf{OS}(\tau_{i,j}) + \mathsf{OS}(\tau_{j,t})) \\ + (1-\alpha)(R_i.BS + \mathsf{BS}(\tau_{i,j}) + \mathsf{BS}(\tau_{j,t})) \quad (1)$$

When $\alpha = 0$, we select a node only based on the budget score, i.e., selecting the node such that the budget score of the corresponding partial route plus the best budget score from the node to the target node $v_t$ is the smallest. When $\alpha = 1$, the algorithm finds a node such that the objective score of the corresponding partial route plus the best objective score from the node to $v_t$ is minimized.

**Algorithm 3: Greedy Algorithm**

1. $nodeSet \leftarrow \emptyset$; $wordSet \leftarrow \mathcal{Q}.\Omega \setminus v_s.\psi$;
2. $v_{pre} \leftarrow v_s$; $OS \leftarrow 0$; $BS \leftarrow 0$;
3. **for** each word $w_t \in wordSet$ **do**
4.     Get the location set $lSet$ containing $w_t$;
5.     $nodeSet \leftarrow nodeSet \bigcup lSet$;
6. **while** $wordSet$ is not empty **do**
7.     $minS \leftarrow argmin_{v_m \in nodeSet}\text{score}(v_m, R_{pre})$;
8.     $OS \leftarrow OS + \text{OS}(\tau_{pre,m})$; $BS \leftarrow BS + \text{BS}(\tau_{pre,m})$;
9.     $v_{pre} \leftarrow v_m$;
10.     $wordSet \leftarrow wordSet \setminus v_m.\psi$;
11.     Remove the locations containing $v_m.\psi$ from $nodeSet$;
12. $OS \leftarrow OS + \text{OS}(\tau_{pre,t})$; $BS \leftarrow BS + \text{BS}(\tau_{pre,t})$;
13. Return the route found with scores $OS$ and $BS$;

The pseudocode is outlined in Algorithm 3. We use $wordSet$ to keep track of the uncovered query keywords and $nodeSet$ to store all the locations containing uncovered query keywords (line 1). $v_{pre}$ denotes the node where the current partial path ends and is initialized as $v_s$. $OS$ and $BS$ are used to store the objective and budget scores and both initialized to 0 (line 2). We utilize inverted file to find locations for $nodeSet$ (lines 3–5). The algorithm terminates when $wordSet$ is empty. While it is not empty, we find the best node according to Equation 1 (line 7), extend the partial route (line 8–9), and update $wordSet$ and $nodeSet$ (lines 10–11). After we exit the loop, we add the last segment from the partial route's last node to the target node $v_t$ to construct the final route and return (lines 12–13).

Algorithm 3 may fail to find a feasible route even if there exits a feasible one. In each step, it selects the next best node. If we find more nodes at each step, the accuracy will be better while the search space becomes much larger. Hence, it is a tradeoff between the accuracy and efficiency. In the experiments, we study the performance of Algorithm 3 when the best 2 nodes are selected at each step. We denote the algorithm selecting one node by Greedy-1, and the algorithm selecting two nodes by Greedy-2. The worst time complexity of Greedy-1 is $O(mn)$ and for Greedy-2 it is $O(2^m n)$, where $m$ is the number of query keywords and $n$ is the number of nodes in the graph.

Algorithm 3 guarantees that the query keywords are always covered while the budget limit may not be satisfied. This is desirable when the query keywords are important to users (e.g., the users do not want to miss any type of locations in their plan). However, if the budget score is very important (e.g., the users cannot overrun their money budget), we modify this algorithm slightly to accommodate the need. We return a route with budget score not exceeding $\Delta$ while the query keywords may not be totally covered. We break the while-loop when the current partial route cannot be extended any more. That is, in line 6 in Algorithm 3, we check if $L.BS + \text{BS}(\sigma_{l,t}) > \Delta$ instead of if $wordSet$ is empty.

### 3.5 Keyword-aware Top-$k$ Optimal Route Search

We further extend the KOR query to the *keyword-aware top-k route* (K$k$R) query. Instead of finding the optimal route defined in KOR, the K$k$R query is to return the top-$k$ routes starting and ending at the given locations such that they have the best objective scores, cover all the query keywords, and satisfy the given budget constraint. We introduce how to modify the OSScaling algorithm and the BucketBound algorithm for solving K$k$R approximately.

It is relatively straightforward to extend the two approximation algorithms OSScaling and BucketBound for processing the K$k$R query. Due to space limitations, we only briefly present the extension. We need to introduce the definition of "$k$-dominate". A label is "$k$-dominated" if at least $k$ labels dominate it. In the pseudocode of OSScaling algorithm, we need to replace "dominate" by "$k$-dominate." Moreover, instead of keeping track of only the current best result, we need to track the current best $k$ results. The budget score of the $k$th best route is used as the upper bound $U$ to prune unnecessary labels. Similarly, in the BucketBound algorithm, we also apply "$k$-dominate". Moreover, instead of returning immediately when we find a feasible route in the bucket containing $R_{OS}$, the algorithm terminates when we find $k$ feasible routes from the non-empty bucket with the smallest order number.

Note that we do not extend the greedy algorithm for solving K$k$R. The greedy approach is not able to guarantee that a feasible route can be found. Therefore, it is meaningless to return $k$ routes using such a method.

## 4. EXPERIMENTAL STUDY

### 4.1 Experimental Settings

**Algorithms.** We study the performance of the following proposed algorithms: the approximation algorithm OSScaling in Section 3.2, the approximation algorithm BucketBound in Section 3.3, and the greedy algorithms in Section 3.4, denoted by Greedy-1 and Greedy-2 corresponding to selecting the top-1 and top-2 best locations, respectively.

Additionally, we also implemented a naive brute-force approach discussed in Section 3.2. However, it is at least 2 orders of magnitude slower than OSScaling and cannot finish after 1 day, and thus is omitted.

**Data and queries.** We use five datasets in our experimental study. The first one is a real-life dataset collected from Flickr [3] using its public API. We collected 1,501,553 geo-tagged photos taken by 30,664 unique users in the region of the New York city in the United States. Each photo is associated with a set of user-annotated tags. The latitude and the longitude of the place where the photo is taken and its taken time are also collected. Following the work [15], we utilize a clustering method to group the photos into locations. We associate each location with tags obtained by aggregating the tags of all photos in that location after removing the noisy tags, such as tags contributed by only one user. Finally, we obtain 5,199 locations and 9,785 tags in total. Each location is associated with a number of photos taken in the location. Next, we sort the photos from the same user according to their taken time. If two consecutive photos are taken at two different places and the taken time gap is less than 1 day, we consider that the user made a trip between the two locations, and we build an edge between them.

On each edge, the Euclidean distance between its two vertices (locations) serves as the budget value. We compute a popularity score for each edge following the idea of the work [4]. The popularity of an edge $(v_i, v_j)$ is estimated as the probability of the edge being visited: $Pr_{i,j} = \frac{Num(v_i,v_j)}{TotalTrips}$, where $Num(v_i, v_j)$ is the number of trips between $v_i$ and $v_j$ and $TotalTrips$ is the total number of trips. The total popularity score of a route $R = (v_0, v_1, ..., v_n)$ is computed as: $\text{PS}(R) = \prod_{i=1}^{n} Pr_{i-1,i}$. However, the popularity score should be maximized. To transform the maximization problem to the minimization problem as defined in KOR, we compute the objective score on each edge $(v_i, v_j)$ as: $o(v_i, v_j) = log(\frac{1}{Pr_{i,j}})$. Therefore, if $\text{OS}(R)$ is minimized, $\text{PS}(R)$ is maximized.

---
[3] http://www.flickr.com/

The other 4 datasets are generated from real data, mainly for scalability experiment. By extracting the subgraph of the New York road network [4], we obtain 4 datasets containing 5,000, 10,000, 15,000, and 20,000 nodes, respectively. Each node is associated with a set of randomly selected tags from the real Flickr dataset. The travel distance is used as the budget score, and we randomly generate the objective score in the range (0,1) on each edge to create the graphs for the four datasets.

We generate 5 query sets for the Flickr dataset, in which the number of keywords are 2, 4, 6, 8, and 10, respectively. The starting and ending locations are selected randomly. Each set comprises 50 queries. Similarly, we also generate 5 query sets for each of the 4 other datasets.

All algorithms were implemented in VC++ and run on an Intel(R) Xeon(R) CPU X5650 @2.66GHz with 4GB RAM.

## 4.2 Experimental Results

### 4.2.1 Efficiency of Different Algorithms

The objective of this set of experiments is to study the efficiency of the proposed algorithms with variation of the number of query keywords and the budget limit $\Delta$ (travel distance). We set the value for the scaling parameter $\epsilon$ in OSScaling and BucketBound at 0.5, the specified parameter $\beta$ at 1.2 for BucketBound, and the default value for $\alpha$ in Greedy at 0.5. We conduct the experiment to study the runtime when varying the value of $\epsilon$ for OSScaling, and the experiment to study the runtime when varying the value of $\beta$ for BucketBound ($\epsilon$=0.5). Note that the runtime of Greedy is not affected by $\alpha$.

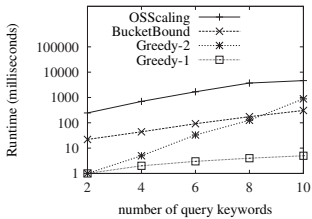
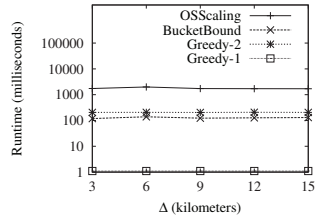

Figure 4: Runtime (Flickr)    Figure 5: Runtime (Flickr)

**Varying the number of query keywords.** Figure 4 shows the runtime of the four algorithms on the Flickr dataset when we vary the number of query keywords. For each number, we report the average runtime over five runs, each using a different $\Delta$, namely 3, 6, 9, 12, and 15 kilometers, respectively. Note that the y-axis is in logarithmic scale. We can see that all the algorithms are reasonably efficient on this dataset. As expected, the algorithm OSScaling runs much slower than the other three algorithms. BucketBound is usually 8-10 times faster than OSScaling, although OSScaling and BucketBound have the same worst time complexity. This is because BucketBound terminates immediately when a feasible route is found in the bucket containing $R_{OS}$, the route found by OSScaling, and thus it generates much fewer labels than does OSScaling. The worst time complexity of both OSScaling and BucketBound is exponential in the number of query keywords. However, as shown in the experiment, the runtime does not increase dramatically as the number of query keywords is increased. This is due to the two optimization strategies employed in both algorithms. Without employing the optimization strategies, both algorithms will be 3-5 times slower. Due to space limitations, we omit the details.

[4]http://www.dis.uniroma1.it/ challenge9/download.shtml

Greedy-1 is the fastest since it only selects the best node in each step. However, as to be shown, its accuracy is the worst. Greedy-1 is not affected significantly by the number of query keywords. The runtime of Greedy-2 increases dramatically with the increase of query keywords. This is because Greedy-2 selects the best 2 nodes at each step, and its asymptotically tight bound complexity is exponential in the number of query keywords.

**Varying the budget limit $\Delta$.** Figure 5 shows the runtime of the four approaches on the Flickr dataset with the variation of $\Delta$. At each $\Delta$, the average runtime is reported over 5 runs, each with a different number of query keywords from 2 to 10. The runtime of OSScaling grows when $\Delta$ increases from 3 km to 6 km as a smaller $\Delta$ can prune more routes. However, as $\Delta$ continues to increase, the runtime decreases slightly. This is due to the fact that with a larger $\Delta$, OSScaling finds a feasible solution earlier (since $\Delta$ is more likely to be satisfied), and then the feasible solution can be used to prune the subsequent search space. The saving dominates the extra cost incurred by using larger $\Delta$ (notice that larger $\Delta$ deteriorates the worst-case performance rather than the average performance). As for the other approximation algorithms, their runtime is almost not affected by the budget limit as shown in the figure.

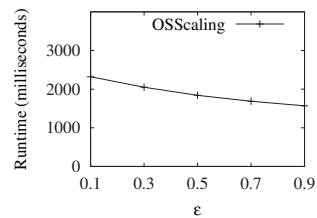
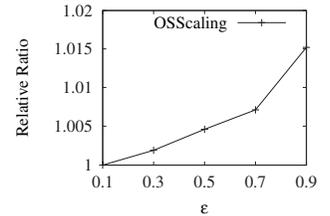

Figure 6: Runtime    Figure 7: Relative Ratio

**Varying the parameter $\epsilon$ for OSScaling.** Figure 6 shows the runtime of OSScaling when we vary the value of $\epsilon$. We set $\Delta$ as 6 km and the number of query keywords as 6. It is observed that OSScaling runs faster as the value of $\epsilon$ increases. This is because when $\epsilon$ becomes larger, $L_{max}$, the upper bound of the number of labels on a node is decreased, and thus more labels (representing partial routes) can be pruned during the algorithm. This is consistent with the complexity analysis of OSScaling, which shows that OSScaling runs linearly in $\frac{1}{\epsilon}$.

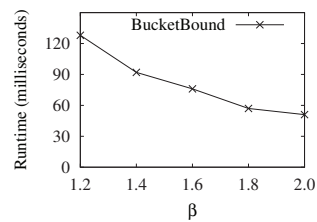
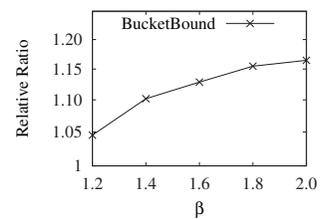

Figure 8: Runtime    Figure 9: Relative Ratio

**Varying the parameter $\beta$ for BucketBound.** Figure 8 shows the runtime of BucketBound when we vary the value of $\beta$, the specified parameter. In this set of experiments, $\Delta$=6 km, $\epsilon$=0.5, and the number of query keywords is 6. As expected, BucketBound runs faster as the value of $\beta$ increases. This is because when $\beta$ becomes larger, the interval of each bucket becomes larger and each bucket can accommodate more labels. Hence, it is faster for BucketBound to find a feasible solution in the bucket containing the best route in $\mathcal{G}$.

### 4.2.2 Accuracy of Approximation Algorithms

The purpose of this set of experiments is to study the accuracy of the approximation algorithms. The brute-forth method discussed in Section 3.2 failed to finish for most of settings after more than 1 day. We note that in the very few successful cases (small $\Delta$ and keywords), the practical approximation ratios of OSScaling and BucketBound are a lot smaller than their theoretical bounds, compared with the exact results by the brute-forth method,. To make the experiments tractable, we study the relative approximation ratio. We use the result of OSScaling with $\epsilon$=0.1 (which has the smallest approximation ratio in the proposed methods) as the base and compare the relative performance of the other algorithms with it. We compute the *relative ratio* of an algorithm over OSScaling with $\epsilon$=0.1 as follows: For each query, we compute the ratio of the objective score of the route found by the algorithm to the score of the route found by OSScaling with $\epsilon$=0.1, and the average ratio over all queries is finally reported as the measure.

With the measure, we study the effect of the following parameters on accuracy, namely the number of query keywords, the budget limit $\Delta$, the scaling parameter $\epsilon$ in OSScaling, the specified parameter $\beta$ in BucketBound, and the parameter $\alpha$ which balances the importance of the objective and budget scores during the node selection, for Greedy.

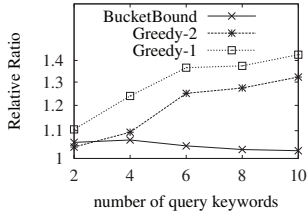 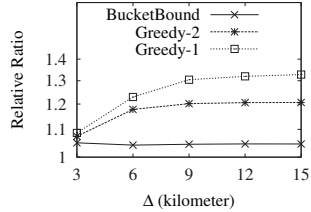

Figure 10: Relative Ratio    Figure 11: Relative Ratio

**Varying the number of query keywords or $\Delta$.** Figure 10 shows the relative ratio compared with the results of OSScaling with $\epsilon$=0.1 for the experiment in Figure 4, in which we vary the number of query keywords. Figure 11 shows the relative ratio for the experiment in Figure 5, in which we vary the value of budget limit $\Delta$, respectively. Note that $\epsilon$=0.5 and $\beta$=1.2 in the two experiments.

Since the greedy algorithms fail to find a feasible solution on about 10%–20% queries, for greedy algorithms we measure the relative ratio only on the queries where Greedy-1 and Greedy-2 are able to find feasible routes. For OSScaling and BucketBound, the reported results are based on all queries, which are similar to the results if we only use the set of queries for which Greedy returns feasible solutions. We observe that the relative ratio of BucketBound compared with the results of OSScaling is always below the specified parameter $\beta$. It can also be observed that BucketBound can achieve much better accuracy than do Greedy-1 and Greedy-2, especially when the number of query keywords or the value of $\Delta$ is large.

**Varying the parameter $\epsilon$ for OSScaling.** Figure 7 shows the effect of $\epsilon$ on the relative ratio in OSScaling. We set $\Delta$ as 6 kilometers and the number of query keywords at 6. We can observe that the relative ratio becomes worse as we increase $\epsilon$, which is consistent with the result of Theorem 2, i.e., the performance bound of OSScaling is $\frac{1}{1-\epsilon}$.

**Varying the parameter $\beta$ for BucketBound.** Figure 9 shows the effect of $\beta$ on the relative ratio in BucketBound, while the corresponding runtime is reported in Figure 8, where we set $\epsilon$=0.5, $\Delta$=6 km, and the number of query keywords as 6. As expected, the relative ratio becomes worse as we increase $\beta$. Note that relative ratio of BucketBound compared to the results of OSScaling is consistently smaller than the specified $\beta$.

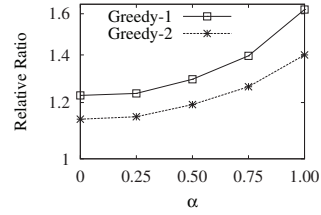 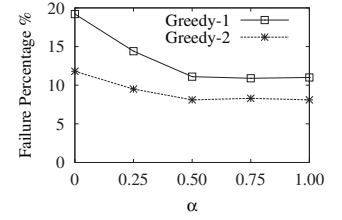

Figure 12: Relative Ratio    Figure 13: Failure Percentage

**Varying the parameter $\alpha$ for Greedy.** Figure 12 shows the relative ratio of Greedy-1 and Greedy-2 compared with the results of OSScaling when we vary $\alpha$, and Figure 13 shows the percentage of failed queries. In this set of experiments, we set $\Delta$ as 6 kilometers, and the average performance is reported over 5 runs, each with a different number of query keywords from 2 to 10. Note that the relative ratio is computed based on the set of queries where Greedy-1 and Greedy-2 are able to find feasible routes over the set of queries with feasible solutions (OSScaling and BucketBound guarantee to return feasible results if any). We observe that as the value of $\alpha$ increases the relative ratio becomes worse for both Greedy-1 and Greedy-2, but they succeed in finding feasible routes for more queries. When $\alpha$ is set as 0, which means that the objective score is the only criterion when selecting the node in each step of Greedy, both Greedy-1 and Greedy-2 achieve the best average ratio while the failure percentage is the largest. When $\alpha$=1, the next best node is selected merely based on the budget score. Hence, Greedy is able to find feasible results on more queries, but the relative accuracy becomes much worse on the queries for which Greedy is able to return feasible solutions. Greedy-2 outperforms Greedy-1 consistently, because more routes are checked in Greedy and it is likely to find more feasible and better routes.

### 4.2.3 Comparing OSScaling and BucketBound

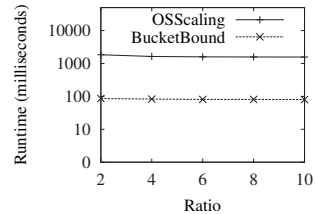 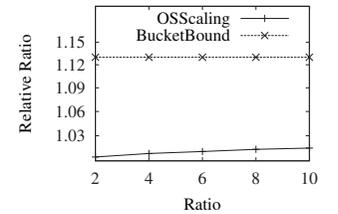

Figure 14: Runtime    Figure 15: Relative Ratio

The aim of this set of experiment is to compare the performance of OSScaling and BucketBound when they have the same theoretical approximation ratio. In this set of experiments, $\Delta$=6 km, $\beta$=1.2, and the number of query keywords is 6. The values of $\epsilon$ are computed according to different performance bounds for both algorithms. Figures 14 and 15 show the runtime and relative ratio of OSScaling and BucketBound when we vary the performance bound, respectively. We observe that BucketBound runs consistently faster than OSScaling over all performance bounds while OSScaling always achieves better relative ratio.

### 4.2.4 Performance of Algorithms for K$k$R

We study the performance of the modified versions of the two approximation algorithms, i.e., OSScaling and BucketBound for processing K$k$R. We set $\epsilon$=0.5, $\beta$=1.2, $\Delta$=6 km, and the average runtime is reported over 5 runs, each with a different number of query keywords from 2 to 10. The results are shown in Figure 16. BucketBound always outperforms OSScaling in terms of runtime. As expected, both algorithms run slower as we increase the value of $k$. In OSScaling, more labels need to be generated for larger $k$, which leads to longer runtime. Algorithm BucketBound terminates only after the top-$k$ feasible routes are found, thus needing longer query time.

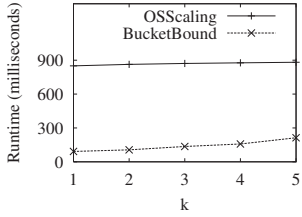
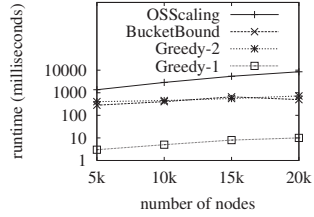

Figure 16: Runtime  Figure 17: Scalability

*4.2.5 Experiments on More Datasets*

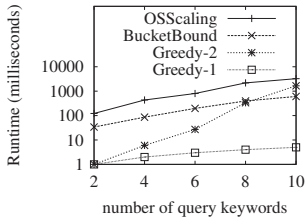
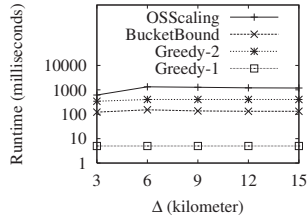

Figure 18: Runtime  Figure 19: Runtime

We also conduct experiments on the synthetic dataset containing 5,000 nodes. Figure 18 and 19 show the runtime when we vary the number of query keywords and the value of $\Delta$, respectively. We set $\epsilon$ as 0.5 and $\beta$ as 1.2. The comparison results are consistent with those on the Flickr dataset. For the relative ratio, we observe qualitatively similar results on this dataset as we do on Flickr. We omit the results due to space limitations.

*4.2.6 Scalability*

Figure 17 shows the runtime of the proposed algorithms (the number of query keywords is 6 and $\Delta$=30 km). They all scale well with the size of the dataset. The relative ratio changes only slightly; we omit the details due to the space limitation.

*4.2.7 Example*

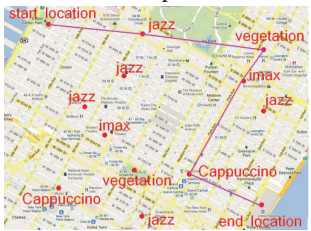
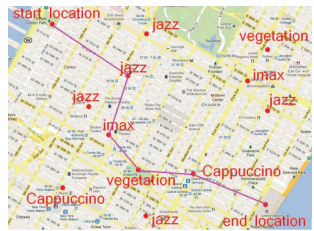

Figure 20: Example Route 1  Figure 21: Example Route 2

We use one example found in the Flickr dataset to show that KOR is able to find routes according to users' various preferences. We set the starting location at the Dewitt Clinton park and the destination at United Nations Headquarters, and the query keywords are "jazz", "imax", "vegetation", and "Cappuccino", i.e., a user would like to find a route such that he can listen to jazz music, watch a movie, eat vegetarian food and have a cup of Cappuccino. When we set the distance threshold $\Delta$ as 9 km, the route shown in Figure 20 is returned by OSScaling as the most popular route that covers all query keywords and satisfies distance threshold. We find that according to the historical trips, this route has the most visitors among all routes covering all the query keywords shorter than 9 km. However, when $\Delta$ is set as 6 km, the route shown in Figure 21 is returned. This route has the most visitors among all feasible routes given $\Delta$=6 km. In the case, the route in Figure 20 exceeds the limit $\Delta$=6 km and is pruned during the execution of OSScaling algorithm.

## 5. RELATED WORK

**Travel route search:** The travel route search problem has received a lot of attention. Li et al. [17] propose a new query called Trip Planning Query (TPQ) in spatial databases, in which each spatial object has a location and a category, and the objects are indexed by an R-tree. A TPQ has three components: a start location $s$, an end location $t$, and a set of categories $C$, and it is to find the shortest route that starts at $s$, passes through at least one object from each category in $C$ and ends at $t$. It is shown that TPQ can be reduced from the Traveling Salesman problem, which is NP-hard. Based on the triangle inequality property of metric space, two approximation algorithms including a greedy algorithm and an integer programming algorithm are proposed. Compared with TPQ, KOR studied in this paper includes an additional constraint (the budget constraint), and thus is more expressive. The algorithms in the work [17] cannot be used to process KOR.

Sharifzadeh et al. [22] study a variant problem of TPQ [17], called optimal sequenced route query (OSR). In OSR, a total order on the categories $C$ is imposed and only the starting location $s$ is specified. The authors propose two elegant exact algorithms L-LORD and R-LORD. Under the same setting [17] that objects are stored in spatial databases and indexed by an R-tree, metric space based pruning strategies are developed in the two exact algorithms.

Chen et al. [3] considers the multi-rule partial sequenced route (MRPSR) query, which is a unified query of TPQ and OSR. Three heuristic algorithms are proposed to answer MRPSR. KOR is different from OSR and MRPSR and the their algorithms are not applicable to process KOR.

Kanza et al. [14] consider a different route search query on the spatial database: the length of the route should be smaller than a specified threshold while the total text relevance of this route is maximized. Greedy algorithm is proposed without guaranteeing to find a feasible route. Their subsequent work [12] develops several heuristic algorithms for answering a similar query in an interactive way. After visiting each object, the user provides feedback on whether the object satisfies the query, and the feedback is considered when computing the next object to be visited. In the work [16], approximate algorithms for solving OSR [22] in the presence of order constraints in an interactive way are developed. Kanza et al. also study the problem of searching optimal sequenced route in probabilistic spatial database [13]. Lu et al. [18] consider the same query [14] and propose a data mining-based approach. The queries considered in these works are different from KOR and these algorithms cannot be used to answer KOR.

Malviya et al. [20] tackle the problem of answering continuous route planning queries over a road network. The route planning [20] aims to find the shortest path in the presence of updates to the delay estimates. Roy et al. [21] consider the problem of interactive trip planning, in which the users give feedbacks for the already suggested points-of-interests, and the itineraries are constructed iteratively based on the users' preferences and time budget. Obviously, these two problems are different with KOR.

Yao et al. [26] propose the multi-approximate-keyword routing (MARK) query. A MARK query is specified by a starting and an ending location, and a set of (*keyword*, *threshold*) value pairs. It searches for the route with the shortest length such that it covers at least one matching object per keyword with the similarity larger than the corresponding threshold value. Obviously, MARK has different aims with that of the KOR query.

The collective spatial keyword search [2] is related to our problem, where a group of objects that are close to a query point and collectively cover a set of a set of query keywords are returned as the result. However, the KOR query requires a route satisfying a budget constraint rather than a set of independent locations. Our problem is also relevant to the spatial keyword search queries [1, 6] where both spatial and textual features are taken into account during the query processing. However, they retrieve single objects while the KOR query finds a route.

**Travel route recommendation:** Recent works on travel route recommendation aim to recommend routes to users based on users' travel histories. Lu et al. [19] collect geo-tagged photos from Flickr and build travel routes from them. They define popularity scores on each location and each trip, and recommend a route that has the largest popularity score within a travel duration in the whole dataset for a city. The recommendation in this work is not formulated as queries and the recommendation algorithm runs in an extreme long time. The work [4] finds popular routes from users' historical trajectories. The popularity score is defined as the probability from the source location to the target location estimated using the absorbing Markov model based on the trajectories. Yoon et al. [27] propose a smart recommendation, based on multiple user-generated GPS trajectories, to efficiently find itineraries. The work [15] predicts the subsequent routes according to the user's current trajectory and previous trajectory history. None of these proposals takes into account the keywords as we do in this work.

## 6. CONCLUSION AND FUTURE WORK

In this paper, we define the problem of keyword-aware optimal route query, denoted by KOR, which is to find an optimal route such that it covers a set of user-specified keywords, a specified budget constraint is satisfied, and the objective score of the route is optimized. The problem of answering KOR queries is NP-hard. We devise two approximation algorithms, i.e., OSScaling and BucketBound with provable approximation bounds for this problem. We also design a greedy approximation algorithm. Results of empirical studies show that all the proposed algorithms are capable of answering KOR queries efficiently, while the algorithms BucketBound and Greedy run faster. We also study the accuracy of approximation algorithms.

In the future work, we would like to improve the current pre-processing approach. We can employ a graph partition algorithm to divide a large graph into several subgraphs. Next, we only do the pre-processing within each subgraph instead of on the whole graph. We also compute and store the best objective and budget score between every pair of border nodes. Thus, the path with the best objective or budget score can be obtained from the pre-processing results. We believe that this approach can greatly reduce the time and space costs of the pre-processing.

## 7. ACKNOWLEDGEMENTS

This work is supported in part by a grant awarded by a Singapore MOE AcRF Tier 1 Grant (RG16/10).